\documentclass[12pt,preprint]{aastex}

\newcommand{\eg}{{\it e.g.}}
\newcommand{\ie}{{\it i.e.}}
\newcommand{\viz}{{\it viz.}}
\newcommand{\etal}{{\it et al.}}

\newcommand{\Msun}{\ensuremath{M_{\odot}}}

\slugcomment{GEMS Neutron Star White Paper}

\shorttitle{GEMS Neutron Star White Paper}
\shortauthors{GEMS SWG}

\begin{document}


\title{White Paper on GEMS Study of \\ Polarized X-rays from Neutron Stars}



\author{Pranab Ghosh\altaffilmark{4},
Lorella Angelini\altaffilmark{1},
Matthew Baring\altaffilmark{2},
Wayne Baumgartner\altaffilmark{1},
Kevin Black\altaffilmark{1},
Jessie Dotson\altaffilmark{3},
Alice Harding\altaffilmark{1},
Joanne Hill\altaffilmark{1},
Keith Jahoda\altaffilmark{1},
Phillip Kaaret\altaffilmark{5},
Tim Kallman\altaffilmark{1},
Henric Krawczynski\altaffilmark{6},
Julian Krolik\altaffilmark{7},
Dong Lai\altaffilmark{8},
Craig Markwardt\altaffilmark{1},
Herman Marshall\altaffilmark{9},
Jeffrey Martoff\altaffilmark{10},
Robin Morris\altaffilmark{3},
Takashi Okajima\altaffilmark{1},
Robert Petre\altaffilmark{1},
Juri Poutanen\altaffilmark{11},
Stephen Reynolds\altaffilmark{12},
Jeffrey Scargle\altaffilmark{3},
Jeremy Schnittman\altaffilmark{1},
Peter Serlemitsos\altaffilmark{1},
Yang Soong\altaffilmark{1},
Tod Strohmayer\altaffilmark{1},
Jean Swank\altaffilmark{1},
Y. Tawara\altaffilmark{13},
and Toru Tamagawa\altaffilmark{14}}

\altaffiltext{1}{NASA/GSFC}
\altaffiltext{2}{Department of  Physics and Astronomy, Rice Univ.}
\altaffiltext{3}{NASA/ARC}
\altaffiltext{4}{Department of Astronomy and Astrophysics, TIFR, Mumbai, India}
\altaffiltext{5}{Department of Physics and Astronomy, University of Iowa}
\altaffiltext{6}{Department of Physics, Washington U. }
\altaffiltext{7}{Department of Physics and Astronomy, Johns Hopkins University}
\altaffiltext{8}{Department of Astronomy, Cornell University}
\altaffiltext{9}{Center for Space Research, MIT}
\altaffiltext{10}{Department of Physics, Temple U.}
\altaffiltext{11}{U. Oulu, Finland}
\altaffiltext{12}{Department of Physics and Astronomy, North Carolina State University}
\altaffiltext{13}{Nagoya University, Japan}
\altaffiltext{14}{Riken University, Japan}


\begin{abstract}

We examine the expected X-ray polarization properties of neutron-star 
X-ray sources of various types, \eg, accretion and rotation powered pulsars,
magnetars, and low-mass X-ray binaries. We summarize the model 
calculations leading to these expected properties. We describe how a 
comparison of these with their observed properties, as inferred from 
GEMS data, will probe the essential dynamical, electromagnetic, plasma,
and emission processes in neutron-star binaries, discriminate between 
models of these processes, and constrain model parameters. An exciting 
goal is the first observational demonstration in this context of the 
existence of vacuum resonance, a fundamental quantum electrodynamical 
phenomenon first described in the 1930s.  

\end{abstract}

\section{Introduction}
\label{sec:intro}
 
A major class of X-ray sources that the Gravity and Extreme Magnetism SMEX
(GEMS) would be studying is neutron stars in various astrophysical
situations, operating as accretion powered pulsars (APPs), rotation
powered pulsars (RPPs), magnetars, accretion powered millisecond X-ray 
pulsars (AMXPs), and bright, non-pulsing low-mass X-ray binaries
(LMXBs) \citep{ShaTu,Ghosh}. APPs, AMXPs and LMXBs are binary systems
where a neutron star accretes matter from its companion, and
conversion of the gravitational binding energy of the accreted matter powers X-ray 
emission. RPPs are thought to generate X-rays by one or more of the
mechanisms summarized in Sec.\ref{sec:rpp}, the ultimate source of
energy for this emission being that of the neutron star's rotation.
The majority of RPPs are single, relatively young 
neutron stars, but there is a population of RPPs in binaries, consisting mostly
of old, \emph{recycled} neutron stars with low-mass companions, and also 
a small number of young neutron stars with massive companions. 
Magnetars are believed to generate X-rays by conversion of the
magnetic energy available from the superstrong magnetic fields of the 
neutron stars in them. These objects are usually thought to be single, 
although there have been occasional suggestions of
magnetars in binaries. All of the above classes of neutron-star X-ray sources
exhibit periodic pulses at the rotation period of the neutron star, except for 
LMXBs, for which these pulses are believed to be unobservable or absent for 
reasons summarized in Sec.\ref{sec:lmxb}, although the neutron stars
in these sources are thought to be rotating fast as a result of the 
recycling process which they are undergoing. However, many LMXBs
undergo X-ray bursts, and during the bursts from some of them, 
periodic oscillations \emph{have} been observed at high frequencies,
which are believed to be the spin frequencies (or simple 
multiples/submultiples thereof) of the underlying neutron stars.  
A classic timing property exhibited by many LMXBs is quasiperiodic 
oscillations (QPO) at low, medium, and high frequencies, to which we
return in Sec.\ref{sec:qpo}. 
Correlating the above timing behaviors of the X-ray 
fluxes from these classes of neutron-star sources with those of their 
X-ray polarization properties (amount/degree of polarization and its 
position angle, or, alternatively, the Stokes parameters) will be a 
principal diagnostic available to GEMS for addressing the scientific 
questions listed in Sec.\ref{sec:disc}.
   
The magnetic field of the neutron star is the most important parameter for
categorizing the X-ray polarization properties at the source (\ie, the 
environs of the neutron star), as expected, since the primary cause for
the emission polarized X-rays from APPs, RPPs and magnetars is this 
magnetic field --- directly or indirectly --- as explained below. 
Among the above classes of neutron-star sources, APPs and RPPs are universally
believed to have the ``standard'' surface field-strengths of $\sim 10^{12}$ G 
canonical for neutron stars,  with a rough spread of about 
one order of magnitude around this value.  Magnetars are thought to have 
\emph{superstrong} fields  $\sim 10^{14} - 10^{15}$ G by this
standard, whence the name. By contrast, LMXBs and AMXPs are thought to 
have fields $\sim 10^{8} - 10^{9}$ G, \emph{low} by the above
standards. Such low fields are believed to have been produced by a
reduction of the standard neutron-star field strengths in the process
of recycling which occurs during the long accretion phases in LMXBs 
\citep{vdH92}. 

For APPs and magnetars, it is their magnetic fields that dominate 
both (a) the production processes of polarized X-rays at or near the neutron-star
surface, and, (b) the subsequent propagation of these polarized X-rays through the 
magnetosphere surrounding the neutron star (see Secs.\ref{sec:app} and 
\ref{sec:magnet}). This happens because the magnetic field strongly
influences both (a) the differential opacity between the ordinary (O) 
and extraordinary (E) polarization modes, and,
(b) the scattering cross-sections \citep{MN85a,MN85b}. For RPPs, the X-ray 
emission process in the magnetosphere is dominated by the local magnetic field, 
and so therefore is the polarization of the resultant radiation (Sec.\ref{sec:rpp}).  
For LMXBs and AMXPs, by contrast, the magnetic field is so low that
its effects are thought to be negligible in the context of X-ray
polarization, so that the latter is described in terms of non-magnetic Compton 
scattering (see Secs.\ref{sec:amxp} and \ref{sec:lmxb}).

\begin{figure}
\epsscale{0.7}
\plotone{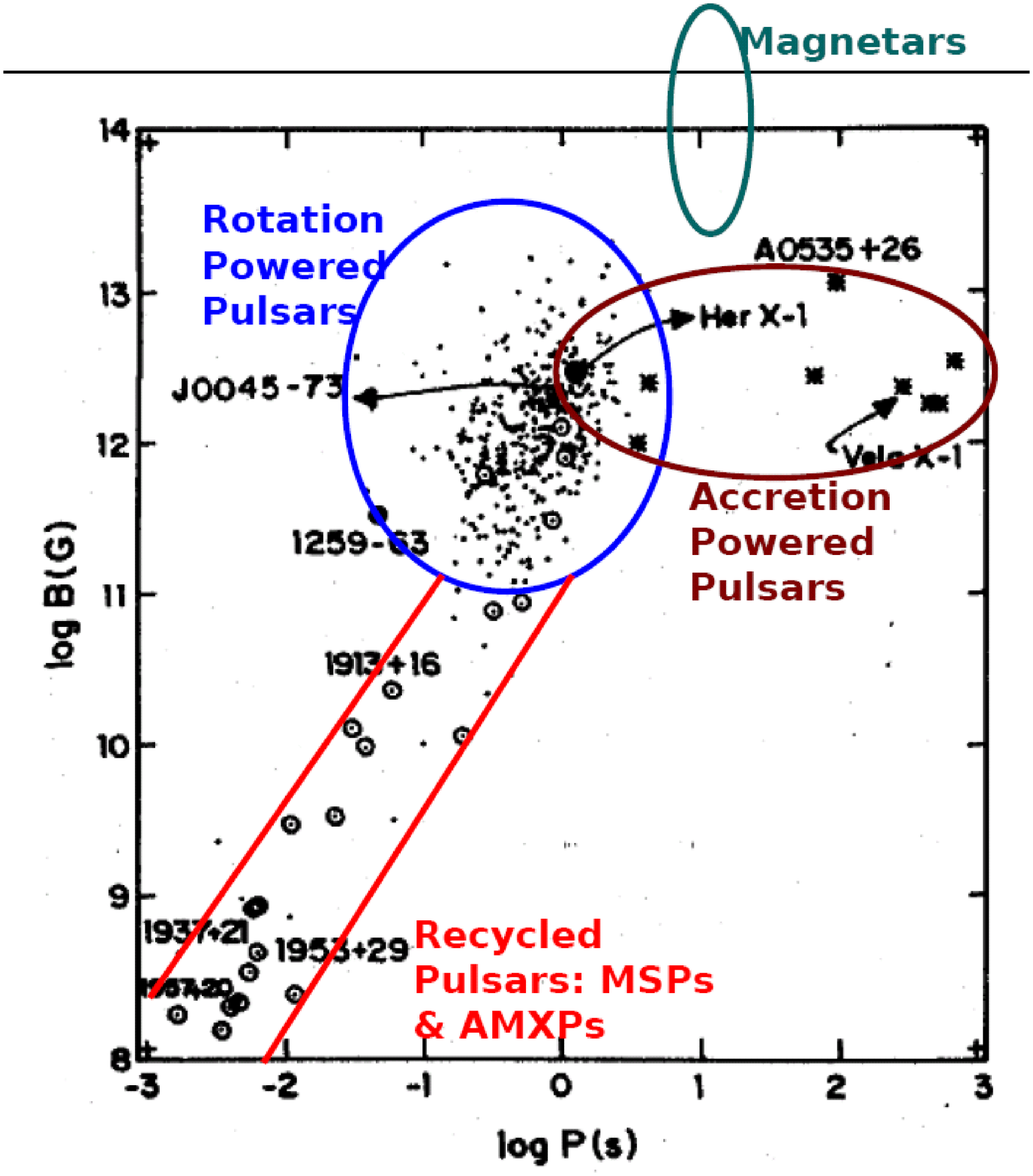}
\caption{Period - Magnetic Field ($P$ - $B$) diagram of neutron stars, 
showing regions containing accretion powered pulsars (APPs, examples
denoted by asterisks), rotation powered pulsars (RPPs, examples
denoted by dots, those in binaries encircled), and magnetars. Recycled
RPPs include both millisecond radio pulsars (MSPs) and accreting 
millisecond X-ray pulsars (AMXPs). Several well-known binary systems 
and single recycled pulsars are labeled. After Ghosh 2007.
\label{fig1}}
\end{figure}

In addition to its magnetic field, the other essential property which 
characterizes a neutron star is its spin period, so that it is most
useful to display neutron stars in a period - magnetic field 
($P - B$) diagram, as shown in Fig.\ref{fig1}. The different classes of
neutron-star systems introduced above occupy characteristically different areas in
this diagram, as shown. The dominant X-ray polarization-producing mechanisms
are also characteristically different. For APPs and magnetars, it is the strong 
differential opacity between O and E modes, while for RPPs, it is the inherent
strongly-polarized nature of the synchrotron/curvature radiation emitted from their 
magnetospheres, and in both cases the polarization degree can be very high, upto  
$\sim 80\%$. For LMXBs and AMXPs, it is the characteristically low polarization
($\sim 10\%-15\%$) signature given to the radiation by non-magnetic Compton 
scattering in the accreting plasma around the neutron star.  

In the following sections, we summarize the current state of knowledge 
of the expected X-ray polarization properties of the above classes of
neutron-star sources, and GEMS strategies for observing and utilizing 
them for diagnostics of emission geometry, magnetic field structure, 
and fundamental quantum electrodynamic (QED) effects.  
In Sec.\ref{sec:app}, we consider APPs. RPPs are considered in 
Sec.\ref{sec:rpp}, and magnetars are dealt with
in Sec.\ref{sec:magnet}. Low magnetic field systems are considered in 
later sections, Sec.\ref{sec:amxp} dealing with AMXPs and
Sec.\ref{sec:lmxb} describing LMXBs, including both burst oscillations
and QPO sources. Finally, we summarize our conclusions in
Sec.\ref{sec:disc}, discussing the key scientific questions to be 
addressed by GEMS.

\section{Accretion Powered Pulsars (APPs)}
\label{sec:app}

Plasma accreting onto the neutron star in an APP is channeled by the
stellar magnetic field to the magnetic poles and so forms
\emph{accretion columns} above these poles \citep{Lambet1, DavOs}.
X-rays generated at the bases of the accretion columns are transported through the 
highly magnetized plasma there, and the transport properties (\eg, the scattering and 
absorption cross-sections) depend strongly on the polarization of the X-rays, being
very different for the \emph{ordinary (O)} mode and the \emph{extraordinary (E)} 
mode. When vacuum polarization effects (see below) are negligible, the O mode has 
its electric vector (which is the conventional definition of the 
direction of polarization) lying in the plane defined by the external 
(\ie, stellar) magnetic field {\bf B} and the wave vector 
{\bf k}, while the E mode has its direction of polarization perpendicular to the 
({\bf k} , {\bf B}) plane. When vacuum polarization effects are significant, the mode 
characteristics are more complicated \citep{Petal80}.  This strong disparity between 
the transport of O and E modes leads to a strong polarization of the emergent radiation. 
A simple, rule-of-thumb argument widely used to illustrate the point is that, at
photon energies $E = h\nu$ far below the cyclotron energy $E_c = \hbar(eB/m_ec)
\approx 11.6 B_{12}$ keV (here, $B_{12}$ is $B$ in units of $10^{12}$ G), the opacity 
$\kappa_E$ of the E mode is drastically reduced compared to that of the opacity
$\kappa_O$ of the O mode, scaling roughly as  $\kappa_E \sim (E/E_c)^2\kappa_O$
\citep{LaiHo}. Consequently, the emergent radiation is dominated by
the E mode, which comes from the deeper and hotter layers of plasma, 
and so is strongly polarized in the direction 
of the E mode.  We shall see below that the actual situation is more complicated, 
but that the general thrust of the above argument is qualitatively
correct.

Emission of X-ray pulses from APPs is conceptually visualized in terms of rotating
X-ray beams which can be characterized as  either \emph{pencil}- or 
\emph{fan}-shaped in two useful limits.  In the former, the opacity
along the accretion column is sufficiently low that the beam emerges preferentially 
along the column, in a pencil shape. In the latter, the opacity along the column is
so high that the beam emerges preferentially through the sides of the column, in a
fan shape \citep{Lambet1}. Pulse profiles for these beam geometries are different 
\citep{MN85b}, and so are the polarization properties \citep{Metal88}. 
We follow here the description of these properties pioneered by
Meszaros and co-authors on the basis of their extensive numerical treatment of 
radiative transfer through the strongly mgnetized plasma in the accretion columns
\citep{MN85a}.

In the Meszaros \etal~approach, the pencil beam is modeled as emission from the
face of a plasma slab lying on the stellar magnetic pole (the stellar magnetic field being 
perpendicular to the slab surface), while the fan beam is modeled as emission from the 
sides of a plasma cylinder sticking out of the stellar surface around the magnetic pole 
(the stellar magnetic field being along the cylinder axis). Transfer of radiation is handled
in either geometry through the Feautrier formalism \citep{feau}, using scattering 
and absorption cross-sections for O and E modes, and a redistribution matrix that 
keeps track of the scattering of photons from one angle to another, from one energy 
to another, and from one polarization mode to the other \citep{MN85a}. 

In a strong magnetic field, not only is the plasma birefringent, but
so also is the vacuum, the latter being a fundamental quantum
electrodynamical (QED) effect first pointed out long ago \citep{HE36}. These two effects
generally work ``against'' each other in a sense, and so ``compensate'' each other 
at a point called \emph{vacuum resonance} \citep{MV79}, where the
degree of linear polarization vanishes\footnote{Actually, the
directions of the dominant linear polarizations generated by 
plasma and vacuum birefringence are perpendicular to each other. At vacuum resonance, 
their strengths are equal, so that the resultant polarization is
circular, not linear.}.  This vacuum resonance energy $E_v$ depends 
on both the magnetic field strength $B$ and the 
electron density $n_e$, and is given by $E_v \approx 13 B_{12}^{-1}n_{e,22}^{1/2}$ keV
(here, $n_{e,22}$ is $n_e$ in units of $10^{22}$ cm$^{-3}$, and $B_{12}$ as before). 
For photon energies below $E_v$, the normal modes of the plasma dominate in the 
radiation-transfer process, while those of the vacuum dominate at $E>E_v$. 
Cross-sections for scattering and absorption for both E and O modes show strong 
features both at the cyclotron energy $E_c$ and at the vacuum
resonance energy $E_v$ \citep{MN85a}, so that the polarization
properties of the emergent X-rays from APPs go through 
strong changes when the X-ray photon energy passes through either of these 
resonances, providing valuable diagnostics of magnetic fields and plasma conditions.
For canonical field strengths of APPs (see above), it is clear that
$E_c$ would be outside the energy range of GEMS operation, 
so that our focus here will necessarily be on the expected changes in 
the essential polarization parameters at $E_v$. 

\begin{figure}
\epsscale{0.7}
\plotone{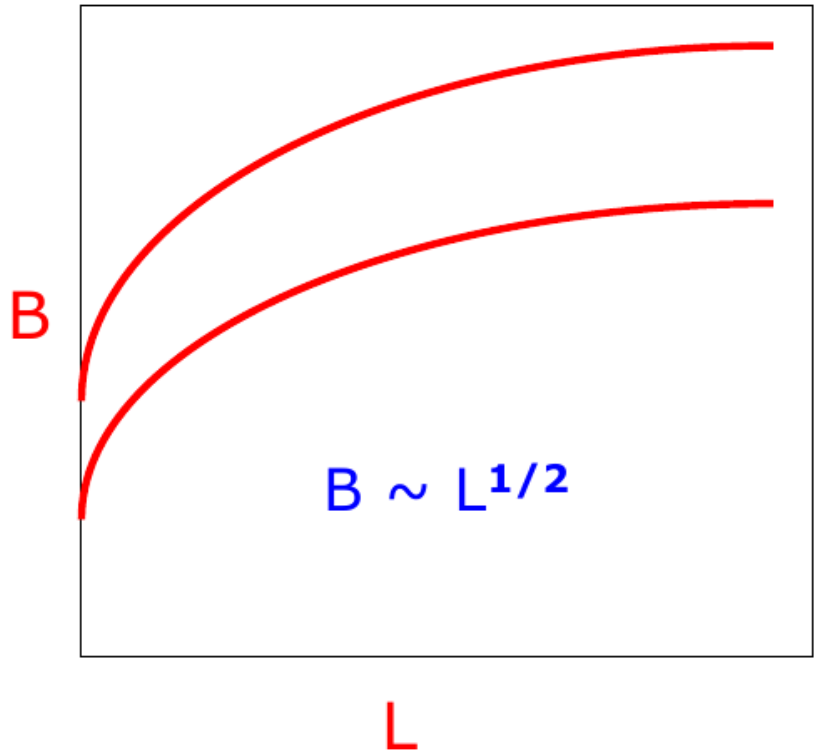}
\caption{Allowed region, between the two solid lines, in the $B-L$
plane, where the vacuum resonance energy $E_v$ falls in the GEMS 
X-ray band (see text). Only the schematic form is shown, with a 
rough scaling given by standard accretion theory. The upper line
corresponds to the lower end of the GEMS band, and the lower line
to the upper end.}
\label{fig2a}
\end{figure}

The GEMS energy-band of $\sim 1-10$ keV defines an allowed region in the 
Magnetic Field - Luminosity ($B-L$) plane of APPs, so that, within
this region, $E_v$ lies \emph{inside} the GEMS band, and therefore 
valuable diagnsotics at $E_v$ would be possible with GEMS. A rough 
sketch of this allowed region is shown in Fig.\ref{fig2a}, for canonical 
properties of accreting neutron stars and accretion geometry \citep{GKS}.

\begin{figure}
\epsscale{0.7}
\plotone{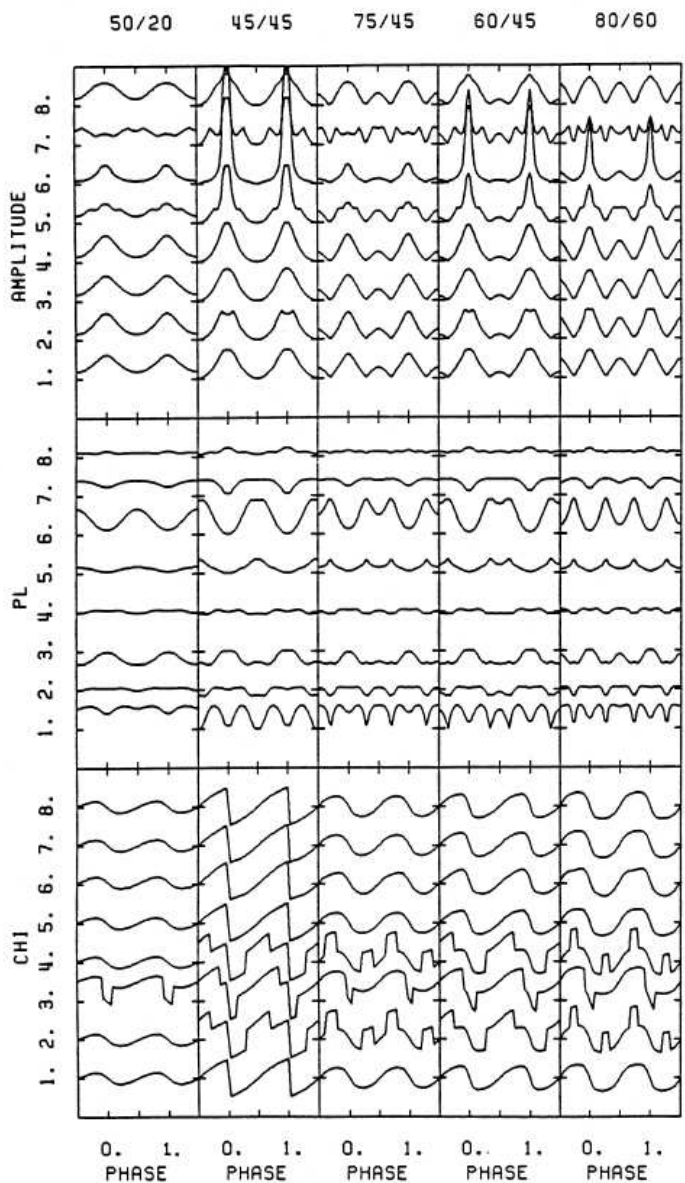}
\caption{Atlas of pulse profiles (top panel), polarization amounts
(middle panel), and position angles (bottom panel) for APPs with
\emph{pencil beams}. 
In each panel, the five columns correspond to different geometries,
labeled by the value of $i_1/i_2$, where $i_1$ = observer angle 
referred to the rotation axis, $i_2$ = inclination angle between 
magnetic and rotation axes, both in degrees. The eight curves in 
each panel and column correspond to different photon energies, 
labels 1 through 8 referring to energies of 1.6, 3.8, 9.0, 18.4, 
29.1, 38.4, 51.7 and 84.7 keV, respectively.  
From Meszaros \etal~1988.\label{fig2b}}
\end{figure}

\begin{figure}
\epsscale{0.7}
\plotone{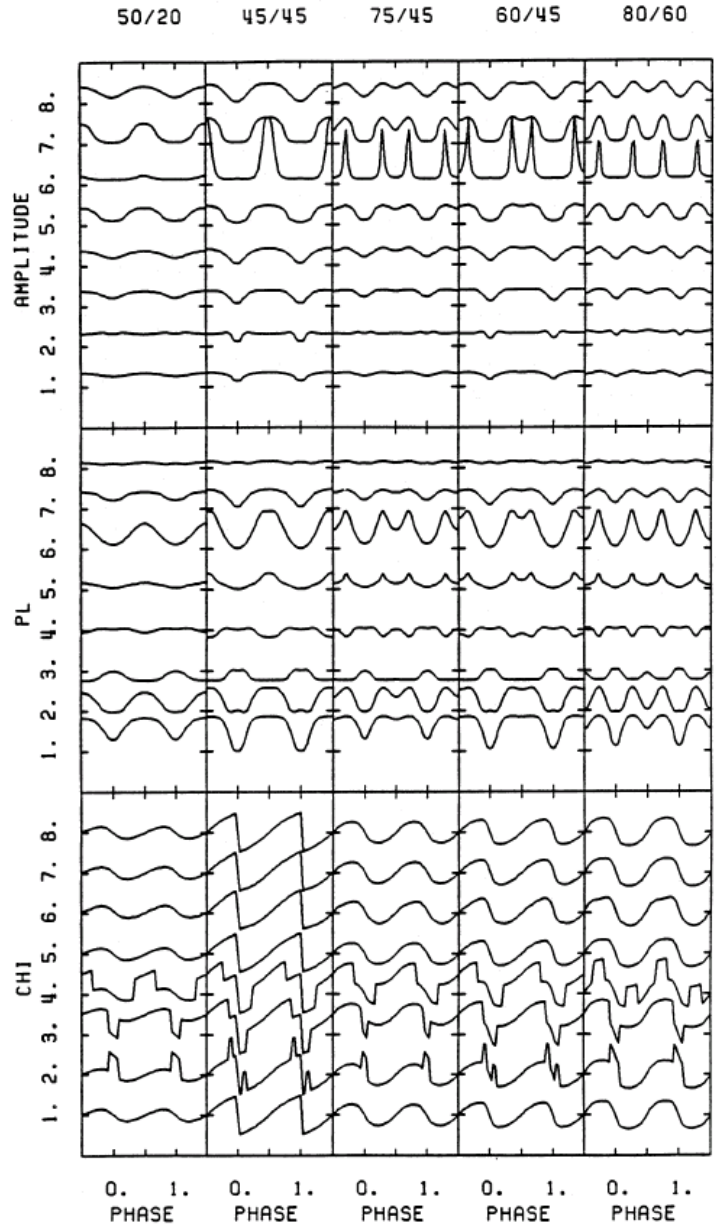}
\caption{Same as Fig.\ref{fig2b}, but for \emph{fan beams}. 
From Meszaros \etal~1988.\label{fig2c}}
\end{figure}

Model X-ray pulse profiles of APPs are shown in Figs.\ref{fig2b} 
and \ref{fig2c}, together with
the (linear) X-ray polarization degree $P_L$ and the polarization
angle $\chi$ as functions of the pulse phase $\phi$, as obtained from 
the numerical scheme of Meszaros \etal, referred to henceforth as
the \emph{M scheme}. For the particular neutron-star and
accreting-plasma parameters used in the results displayed in Figs.\ref{fig2b} 
and \ref{fig2c}, the cyclotron resonance energy is $E_c\approx 38$ keV, and the    
vacuum resonance energy is $E_v\approx 22$ keV. (Note that we have
shown here for illustration the original M scheme results obtained for 
the parameters used by those authors. Recent estimates of these
parameters for well-known, bright APPs indicate that $E_v$ is 
expected to be in the GEMS band for a substantial number of them
\citep{GKS}.) We shall henceforth 
call libraries of such model profiles of total X-ray intensity ($I$), 
$P_L$, and $\chi$ as functions of essential system parameters 
(\eg, the inclination angle and the observer angle) as
\emph{atlases}. Comparison of such atlases 
with GEMS data on APPs will be a major diagnostic probe in our work, and we
hereby name this probe \emph{pulse phase polarimetry}, which will
indeed be a most useful probe for X-ray pulsars of any kind, as we 
shall see in subsequent sections.
 
A key signature of the beam geometry (pencil or fan) is the phase correlation between 
the pulse profile (which we can also call the intensity profile or $I$-profile)
and the $P_L$-profile, which can be seen clearly in Figs.\ref{fig2b} and \ref{fig2c}. 
For pencil beams, the maximum in $P_L$ 
is generally \emph{in phase} with the pulse maximum for photon energies below 
$E_v$, but generally \emph{out of phase} for photon energies above $E_v$. For fan 
beams, the situation is exactly the opposite. In a similar vein, the phase correlation 
between the pulse profile and the $\chi$-profile is also a good diagnostic. For pencil
beams, $\chi$ generally goes from positive to negative values at the
pulse maximum, and from negative to positive values at the pulse 
minimum\footnote{We exclude here discontinuous jumps $\chi$ by 180
degrees, which occur when the line of sight points directly down at
the magnetic pole at some pulse phase, and there is a 
discontinuous jump in the angle between sky-projected stellar magnetic field vector
and the sky-projected rotation axis.}. Again, for fan beams, the situation is exactly 
the opposite.

It follows from the above that a good pictorial representation of the
$I-P_L$ diagnostic would be a direct $I-P_L$ plot, which can be
obtained from the $I$- and $P_L$-profiles, treating the phase $\phi$
as a parameter. This $I-P_L$ plot would be a closed, ellipse-like, but
more complicated curve, in analogy with the familiar Lissajous figure
which would occur if the above profiles were sinusoids \citep{GKS}.
The major axis of this figure would have a positive slope below $E_v$,
but a negative slope above this energy for pencil beams, so that the 
figure would turn dramatically by about 90 degrees as the observation 
energy passes through $E_v$. For fan beams, the situation would be the
opposite, and the turn would be in the opposite direction
\citep{GKS}. Similar considerations apply to the $I-\chi$ plot.
      
The slab or cylinder shaped emission regions envisaged in the above M 
scheme are typically of radial extents $\sim 1-10^3$ m \citep{MN85a,MN85b}, 
\ie, small compared to the neutron-star radius, and quite tiny compared to 
the typical extent of an APP magnetosphere ($\sim 10^6-10^7$ m). 
Polarized X-rays produced by these regions propagate subsequently
through the magnetospheres of APPs, which lead to possible
modifications of their properties. There are two major propagation 
effects whose polarization signatures on X-rays from APPs need 
to be quantified fully. 

The first effect is the resonant Compton scattering of an outgoing
photon generated by the M scheme from those points in the accretion 
column where the photon's frequency equals the local cyclotron frequency 
in the rest frame of the infalling plasma. In the rest frame of 
the \emph{neutron star}, then, the corresponding condition in terms 
of the energies defined earlier can be written as
\begin{equation}
\label{eq:compres}
E = {E_c \over \gamma(1-\beta\mu)}~.
\end{equation}
Here, $\mu\equiv {\bf \hat{k}.\hat{B}}$ is the cosine of the  angle 
between the photon propagation (${\bf \hat{k}}$) and magnetic field 
(${\bf \hat{B}}$) directions, $\beta\equiv v/c$ is the dimensionless 
electron velocity, and $\gamma\equiv (1-\beta^2)^{-1/2}$ is the
Lorentz factor. There is a limiting surface of last scattering,
or \emph{escape} surface, outside which it is not possible to satisfy 
the condition given by Eq.~(\ref{eq:compres}). The location of this 
escape surface is given by the condition $(E_c/E)^2 + \mu^2 = 1$ 
\citep{FD11}. Its radius $r_{es}$ has been calculated for magnetars 
(see Sec.\ref{sec:magnet}), and the values for APPs with characteristically 
lower magnetic fields (see Sec.\ref{sec:intro}) are correspondingly
lower, being given for a purely dipolar neutron-star magnetic 
field roughly as 
\begin{equation}
{r_{es}\over R_{NS}} \approx 2.3 \left[\left({E\over 1 {\rm keV}}
\right)^{-1}B_{12}\left(1-\mu^2\right)^{-1/2}\right]^{1/3}~.    
\label{eq:resapp}      
\end{equation}

The second effect is that of the polarization freezing radius $r_{pl}$. As the X-ray 
photons propagate outward in the magnetosphere, normal modes do not mix 
significantly at first: this propagation is called \emph{adiabatic}, as each mode 
follows its own evolution in the changing magnetic field without being affected by the 
other one. At a sufficiently large radius $r_{pl}$ and small magnetic field strength, 
however, this adiabaticity breaks down and the modes mix, thereby fixing the 
polarization degree and direction for all further outward propagation. $r_{pl}$ has
been calculated for magnetars (see Sec.\ref{sec:magnet}), 
where the entire magnetosphere is dominated by vacuum polarization
effects \citep{FD11}. This can be seen by calculating the radius 
$r_v$ of the vacuum resonance point (see above), and remembering that vacuum
polarization effects dominate for $r<r_v$, while plasma polarization effects dominate 
for  $r>r_v$. For magentars,  $r_v \approx 3000R_{NS}$, where $R_{NS}$ is the 
neutron-star radius \citep{FD11}. For APPs, on the other hand, 
the magnetosphere is almost entirely dominated by plasma 
polarization effects, since the vacuum resonance radius is 
given by \citep{GKS}
\begin{equation}
\label{eq:vacresapp}
{r_v\over R_{NS}} \approx 0.91 \left({E\over 1 {\rm Kev}}\right)^{4/7}B^{20/49}_{12}
R^{1/49}_6\left({M\over\Msun}\right)^{19/49}L^{-10/49}_{37}~.
\end{equation}    
It is clear from Eq.~(\ref{eq:vacresapp}) that, except at high photon energies, high
magnetic fields, and very low luminosities, the entire magnetosphere is dominated by
plasma polarization effects. Thus, a calculation of $r_{pl}$ for APPs involves 
\emph{only} the properties of plasma polarization modes, and so is entirely different
from that for magentars \citep{GKS}.

A third, interesting, fundamental effect is that of general relativity on photon 
propagation near the neutron-star surface, but it appears that its strength is not
large enough for neutron-star radii given by modern equations of state (EOS) to
warrant a detailed inclusion into a polarization computational scheme at
this stage. Gravitational bending of photon trajectories can lead to 
interesting effects on the \emph{pulse profiles} of APPs which were
studied in the 1980s \citep{RM88,MR88}. Briefly, a fraction of the 
photons emitted near the stellar surface can be so bent as to 
propagate backward, so that this backward flux would partly (a) 
be \emph{blocked} by the stellar surface, and partly (b) 
\emph{enhance} the flux from the other magnetic pole \citep{RM88}. 
This would occur for both pencil and fan beams in the M scheme
introduced above. Consequent changes in the pulse profile, particularly 
in the degree of modulation, were studied by the above authors \citep{MR88}. 
As expected, the strength of the GR effects increases with increasing
\emph{compactness} of the neutron star, \ie, decreasing 
values of the parameter $\alpha\equiv R_{NS}/R_s$, where $R_s = 
2GM_{NS}/c^2$ is the Schwarzschild radius. For $M_{NS} = 1.4\Msun$, $R_s
\approx 4.2$ km. The above authors studied the parameter space 
$1.6\le\alpha\le 4.0$ (which correspond to $R_{NS}$ between 6.7 km and 
16.8 km) and showed that GR effects became important for $\alpha\le
2.0$, which corresponds to $R_{NS}\le 8.4 $ km. Those neutron-star EOS 
which are currently considered viable give values of $R_{NS}$ in a 
narrow range around $\approx 12$ km (corresponding to $\alpha\approx 
3$), where GR effects would be minor, estimated to be $\approx 10\%$ 
or less, for both pulse profiles and polarization properties 
\citep{Metal88}. Accordingly, we shall not consider these effects 
further in this paper.

\section{Rotation Powered Pulsars (RPPs)}
\label{sec:rpp}

Radio polarization measurements have been a valuable probe of
the emission geometry in RPPs, and the only \emph{measurement} (\ie, 
not an upper limit) of X-ray polarization from neutron stars so far 
has been from the well-known RPP, the Crab pulsar, and its nebula 
\citep{Wetal76}. However, the emission sites and mechanisms for 
X-ray and radio emission are believed to be entirely different 
because the X-ray and radio pulse profiles are quite different 
in general. There are three main types of models for X-ray (and 
other high energy) emission from RPPs, namely, (a) polar cap 
models (see Daugherty \& Harding 1996 and references therein), 
where the emission occurs within a few stellar radii of the neutron-star
surface, (b) the outer gap models \citep{Cetal86}, where the emission occurs in 
the outer magnetosphere near the light cylinder, and, (c) striped 
wind models \citep{PK05}, where the emission occurs in the pulsar wind 
outside the light cylinder. In the slot gap model, a recently 
explored variation of polar cap-type models, the emission occurs at
high altitudes along the last open magnetic field line, in the 
outer magnetosphere \citep{MH03}.

Polar-cap models assume that particles begin accelerating near 
the neutron-star surface, and emit high-energy radiation at radii
comparable to $R_{NS}$ through curvature radiation and/or inverse
Compton-induced pair cascade in the strong magnetic fields there.
The slot-gap model is somewhat similar, except that the curvature, 
synchrotron, and inverse Compton components of the radiation 
originate in the outer magnetosphere. Outer-gap models assume 
that the acceleration occurs in vacuum gaps that form between the  
null-charge surfaces and the light cylinder in the outer magnetosphere,
and that the observed high-energy radiation is curvature radiation 
and photon-photon pair-production induced cascades from these
sites. Striped-wind models envisage the pulsed emission as 
originating in the pulsar wind beyond the light cylinder, due to
toroidal magnetic fields of alternating sign in this wind, which can 
form current sheets for pulsars with high inclination, and the
reconnection of these magnetic fields can convert magnetic energy 
into particle energy, leading to sharply (double) peaked pulse profiles
if the dissipation region is small enough. Both slot-gap and outer-gap
models have \emph{caustics} in their emission pattern, \ie, photons 
emitted over a broad range of altitudes tend to \emph{pile up} at 
roughly the same phase, due to an almost complete cancellation 
between (a) the phase delay in photon emission with increasing altitude
along a trailing field line, and, (b) special-relativity effects like the 
aberration of photon-emission directions and the time-of-flight effects 
caused by the finite speed of light. In the \emph{two-pole-caustic 
(TPC)} configuration, strong caustics form on field lines trailing both
polar caps, leading to light curves with widely-separated sharp double
peaks.

\begin{figure}
\epsscale{0.6}
\plotone{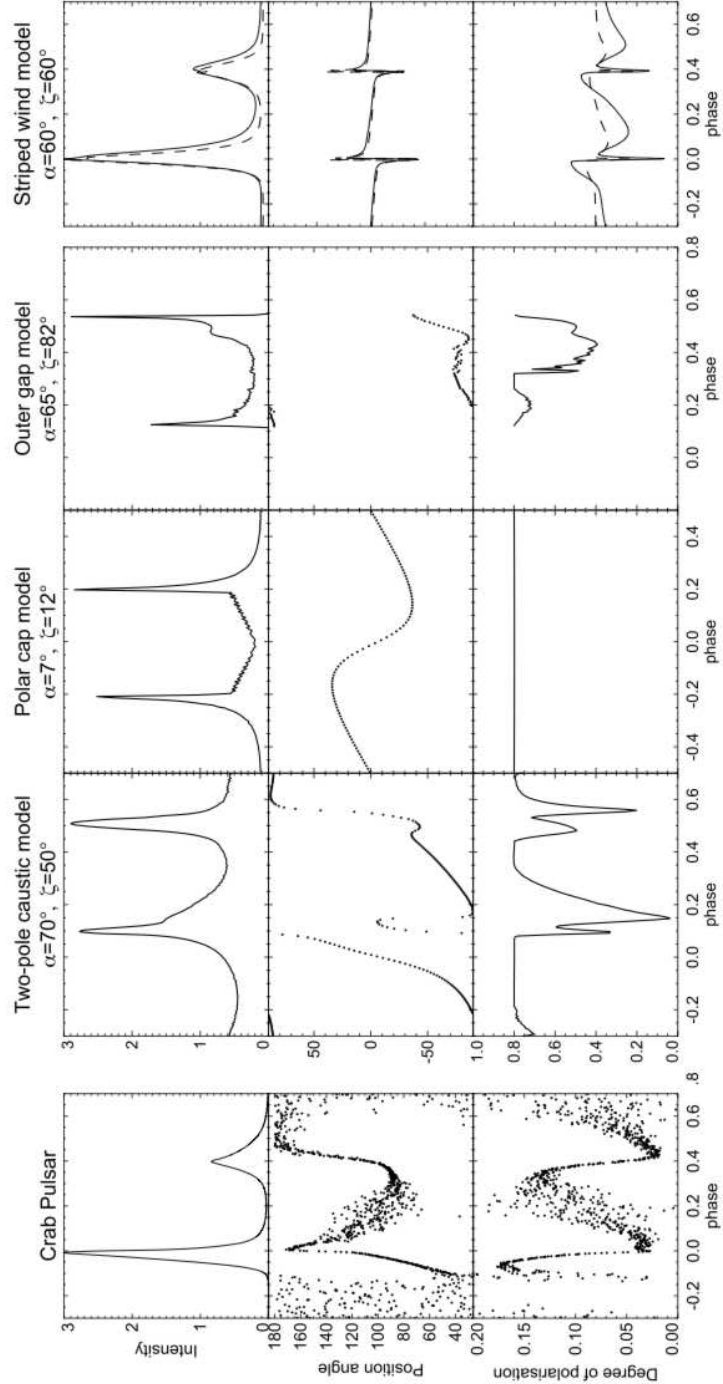}
\caption{Pulse profiles, polarization amounts, and position angles for the Crab pulsar, 
showing optical observations from 
Slowikowska et al. (2009) compared with polar cap, slot gap/two-pole caustic, 
and outer gap models (see text) from
Dyks et al. (2004) and the striped wind model from Petri \& Kirk (2005).  
The latter model is calculated for two Lorentz factors,
20 and 50, shown as solid and dashed lines, respectively.
$\alpha =$ inclination angle between rotation and magnetic axes, 
$\zeta =$ observer angle with respect to the rotation axis.
From Slowikowska et al. (2009).\label{fig3}}
\end{figure}

In the light of extensive studies of high-energy emission from RPPs with
$Fermi$ over the last several years, it is now clear that observations
strongly disfavor polar-cap type models where this emission occurs 
near the neutron-star surface, and strongly support those types of 
models in which this emission takes place in the outer magnetosphere 
or beyond the light cylinder (Harding 2010 and references therein). 
Accordingly, we shall consider only 
the latter types here for X-ray polarization studies, giving an
occasional reference to results for polar-cap type models to show the
difference. Expected pulse-phase variation profiles of the X-ray 
polarization have to be calculated numerically for these
models, since simple, empirical models like the rotating vector 
model, so useful for lower-altitude radio emission (see, \eg, Ghosh
2007 and references therein), are no longer
relevant here. Results from such calculations for the slot-gap and
outer-gap models are shown in Fig.\ref{fig3}, with that for the polar-cap 
model also shown for reference. The latter shows an S-shaped swing
in the polarization angle, reminiscent of the rotating vector model, 
but irrelevant here for the above reasons and also in strong 
disagreement with optical polarization measurements of the Crab 
pulsar. The polarization signatures of the TPC/slot-gap and 
outer-gap models are very different indeed. In particular, for 
the TPC/slot-gap models, the polarization angle shows very rapid
sweeps through the main peak and the interpulse, and a strong
depolarization associated with each peak. These features are
actually observed in the \emph{optical} polarization profile of the 
Crab pulsar \citep{S88}.           
      
\begin{figure}
\epsscale{1.1}
\plotone{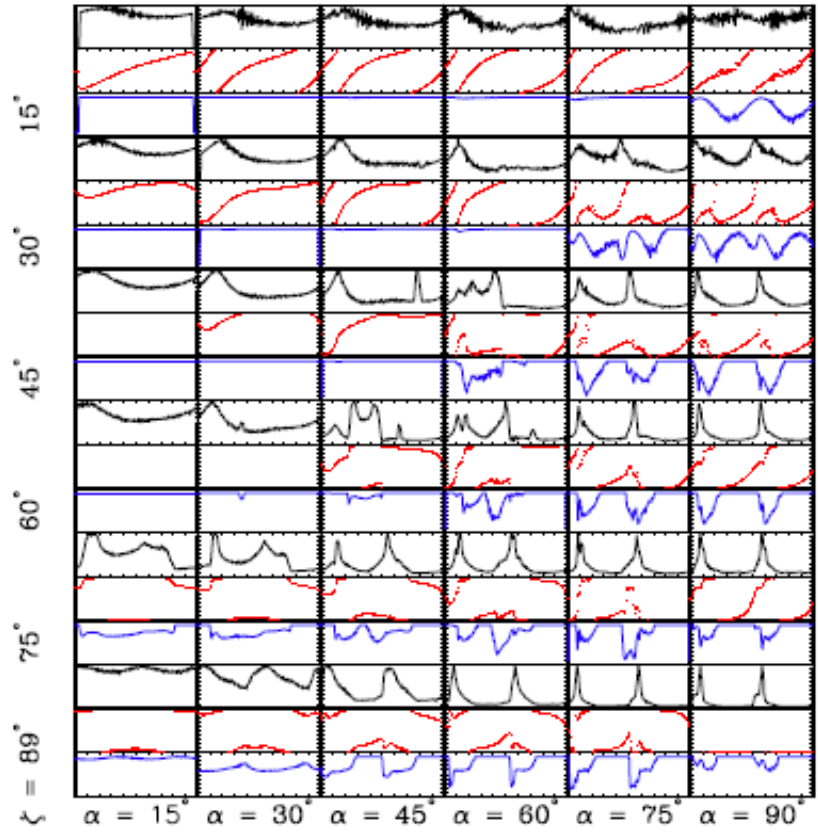}
\caption{Atlas of pulse profiles (black), polarization amounts (blue), and 
position angles (red) for RPPs with a vacuum (Deutsch) magnetic field
configuration in the two-pole caustic (TPC) model (see text). Angles  
$\alpha$ and $\zeta$ as defined in caption of Fig.\ref{fig3}. 
From Harding 2012.\label{fig4a}}
\end{figure}

\begin{figure}
\epsscale{1.1}
\plotone{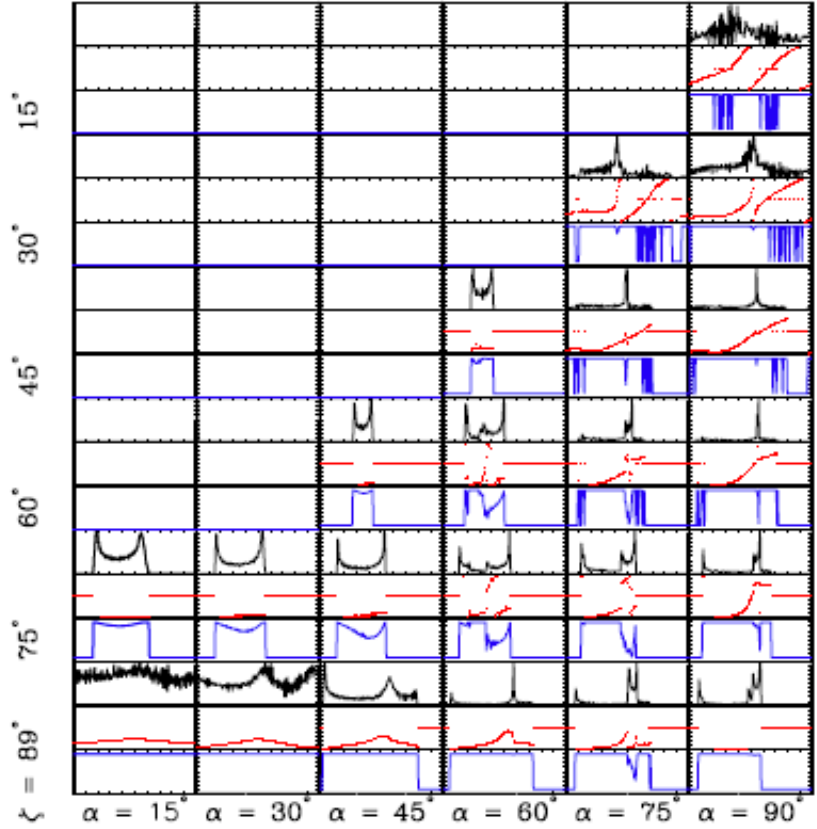}
\caption{Same as Fig.\ref{fig4a}, but for the outer gap (OG) model 
(see text).   
From Harding 2012.\label{fig4b}}
\end{figure}

As the different outer-magnetosphere models produce high
energy RPP pulse profiles which are rather similar, it would be 
difficult to discriminate between them on this count. However,
since different models of this type yield quite different polarization 
profiles at these energies because of their different magnetic
field structures in the outer magnetosphere, pulse-phase  
polarimetry in the X-rays would serve as an excellent discriminator 
between models, and this would be a major direction of RPP study with 
GEMS. To this end, we shall have \emph{atlases} of the profiles 
of $I$, $P_L$, and $\chi$ corresponding to the above models, each such 
atlas detailing the profiles for a particular model for a grid of values 
for the essential parameters, \eg, the inclination angle and the 
observer angle. The idea is exactly analogous to what we described 
in Sec.\ref{sec:app} for APPs, and examples of such RPP atlases are
shown in Figs.\ref{fig4a} and \ref{fig4b}.

\section{Magnetars}
\label{sec:magnet}

\begin{figure}
\epsscale{0.5}
\plotone{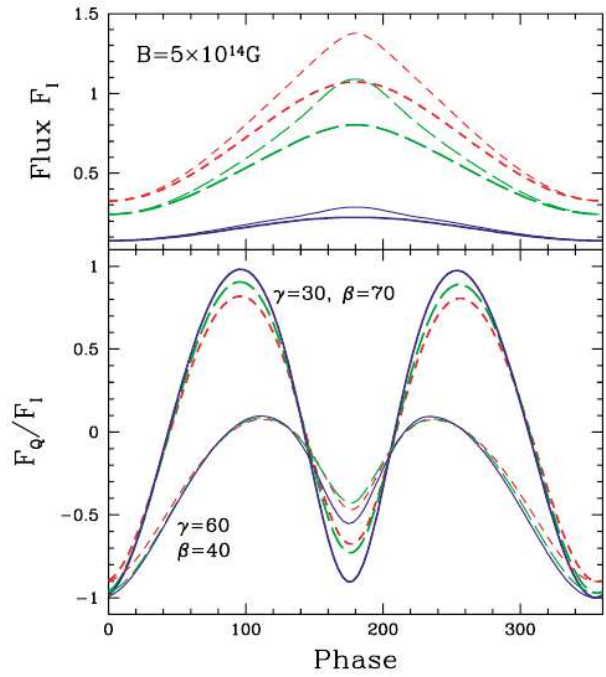}
\caption{Pulse profiles and polarization amounts for magnetar emission from
neutron-star atmospheres. Line style coded by photon energy: 1keV (dashed),
2 keV (long-dashed), 5 keV (solid). $\gamma$ = observer angle referred
to the rotation axis, $\beta$ = inclination angle between rotation and
magnetic axes. 
From Lai and Ho 2003.\label{fig5}}
\end{figure}

Magnetars are believed to generate X-rays by conversion of the magnetic energy 
stored in the twisted magnetospheres of neutron stars with superstrong magnetic
fields (see Sec.\ref{sec:intro}). This energy release near the stellar surface is 
thought to produce ``hot spots'', which emit thermal X-rays through the very thin 
(typically $1-10$ cm thick) neutron-star atmosphere. The polarization properties 
of this primary radiation are determined by the interplay between the radii of the 
photospheres of O and E modes on the one hand, and the vacuum resonance radius 
$r_v$ on the other (see Lai \& Ho 2003 and references therein). At magnetar field 
strengths, the vacuum resonance lies between the O- and E-mode photospheres 
(the latter always lying deeper because of the lower opacity of the E-mode), and 
the emergent radiation from the atmosphere is dominated  by the E mode at all 
photon energies of interest for GEMS observations. Expected pulse-phase 
variations in the polarization of this radiation, as given by model calculations 
\citep{LaiHo}, are shown in Fig.\ref{fig5}.  

Propagation of the above polarized X-rays through magnetar magnetospheres has
two major effects on the polarization, which have been already introduced in 
Sec.\ref{sec:app}, \viz, resonant Compton scattering and polarization freezing.
The first is described by the limiting surface of last scattering or the \emph{escape} 
surface, outside which it is not possible to satisfy the condition given by 
Eq.~(\ref{eq:compres}). The location of this escape surface is given by the condition 
$(E_c/E)^2 + \mu^2 = 1$ and its radius $r_{es}$ has been calculated \citep{FD11} 
for the standard twisted-dipole magnetospheric structure \citep{TLK} 
of magnetars to be 
\begin{equation}
{r_{es}\over R_{NS}} \approx 12 \left[\left({E\over 1 {\rm keV}}\right)^{-1}B_{14}
\xi\left(1-\mu^2\right)^{-1/2}\right]^{1/2.88}~.    
\label{eq:resmag}      
\end{equation}
Here, $B_{14}$ is the polar magnetic field of the magnetar in units of $10^{14}$ G,
$\xi$ is a geometrical factor describing the angular dependence of the magnetic
field, and the specific result given by Eq.(\ref{eq:resmag}) is for a twist parameter of
unity \citep{FD11}.  

The second effect, \viz, polarization freezing, is described in terms
of a polarization freezing radius $r_{pl}$, 
the physics of which has been introduced in Sec.\ref{sec:app}. 
Magnetar magnetospheres are completely dominated by vacuum polarization effects,
as seen by calculating the vacuum resonance point radius (see Sec.\ref{sec:app}) 
$r_v \approx 3000R_{NS}$ \citep{FD11}. Using the properties of the vacuum modes, 
$r_{pl}$ for magnetars has been calculated as 
\begin{equation}
{r_{pl}\over R_{NS}} \approx 146 \left[\left({E\over 1 {\rm keV}}\right)^{-1}B^2_{14}
R_6\xi^2\zeta^{-1}\left(1-\mu^2\right)\right]^{1/4.76}~.    
\label{eq:rplmag}      
\end{equation}
Here $\zeta$ is a dimensionless function of order unity involved in
calculating the scale length on which $B$ changes in the magnetosphere.

\begin{figure}
\epsscale{0.5}
\plotone{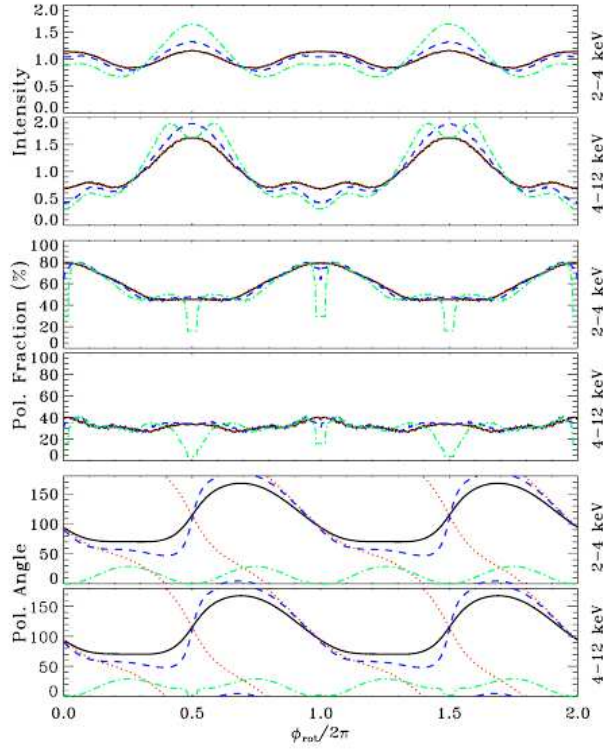}
\caption{Pulse profiles, polarization amounts and position angles for
the final magnetar emission after including propagation effects in the 
magnetosphere. Results shown for two energy bands: 2-4 keV and 4-12
keV. Line style/color coded by geometry, \ie, the combination 
(inclination angle, observer angle), as follows. (45,70): solid/black, 
(70, 45): dotted/red), (60, 70) : dashed/blue, and 
(90, 90) : dot-dashed/green. All angles in degrees.
From Fernandez and Davis 2011.\label{fig6}}
\end{figure}

The above separation by roughly one order of magnitude between 
$R_{NS}$, $r_{es}$, and $r_{pl}$ for magnetars makes the calculation
of propagation effects relatively straightforward for them, since 
resonant scattering is basically decoupled from subsequent
polarization freezing \citep{FD11}, and both are, in a sense, decoupled from 
the processes in the neutron star's atmosphere. Such calculations 
have been done for the propagation of an assumed completely polarized
radiation from the neutron-star atmosphere through a model magnetar
magnetosphere, and the results for the final emergent radiation are
summarized in Fig.\ref{fig6}. The propagation effects tend to reduce the 
amount of polarization, but polarization degrees are still relatively
high ($\sim 40-80 \%$) at lower ($\sim 2-4$ keV) energies, and 
particularly so around the pulse minima. The calculated pulse-phase
variation profiles of the X-ray polarization of magnetars are 
generally similar to those of pencil-beam APPs (described by the 
slab model in the M scheme: see Sec.\ref{sec:app}), as expected, 
since hot-spot emission from neutron-star surfaces in magnetars 
would have a pencil-beam character.

\section{Accretion Powered Millisecond X-ray Pulsars (AMXPs)}
\label{sec:amxp}

In this and the next section, we consider systems containing 
accreting neutron stars with low magnetic fields $\sim 10^8 - 
10^9$ G, where the magnetic field is assumed to have a negligible
effect on radiation transport, and X-ray polarization properties
of such systems are described in terms of Compton scattering in a
plasma configuration with a geometry appropriate for the system. We
consider AMXPs first, of which there are about thirteen currently 
known. These systems are believed to have been produced by the 
\emph{recycling} of neutron stars in low-mass X-ray binaries        
(LMXBs: see next section), where accretion had both (a) spun up the
neutron stars to millisecond periods and (b) reduced their magnetic
fields to the above levels. 

The emission sites of AMXPs are modeled as ``hot spots'' on the 
neutron star surface, in analogy with APPs. Thermal radiation
from these hot spots undergoes Compton scattering in the plasma 
behind the accretion shock, which is modeled as a plane-parallel
slab of optical depth of order unity ($\tau \sim 1$). The basic 
polarization signatures of Compton scattering are well-known 
\citep{ch60,LS82,NP93}. 

For an optically thick  ($\tau >> 1$) slab the direction of 
the linear polarization is preferentially in the 
plane of the slab, and for the classic problem of a semi-infinite 
slab in the Thompson scattering limit, the maximum degree of 
polarization possible is the Chandrasekhar limit $\approx 12\%$ 
\citep{ch60}. Detailed calculations for $\tau = 1$ have been performed, 
which yield typical polarization degrees $\sim 10\% - 15\%$
\citep{PS96}.

\begin{figure}
\epsscale{0.8}
\plotone{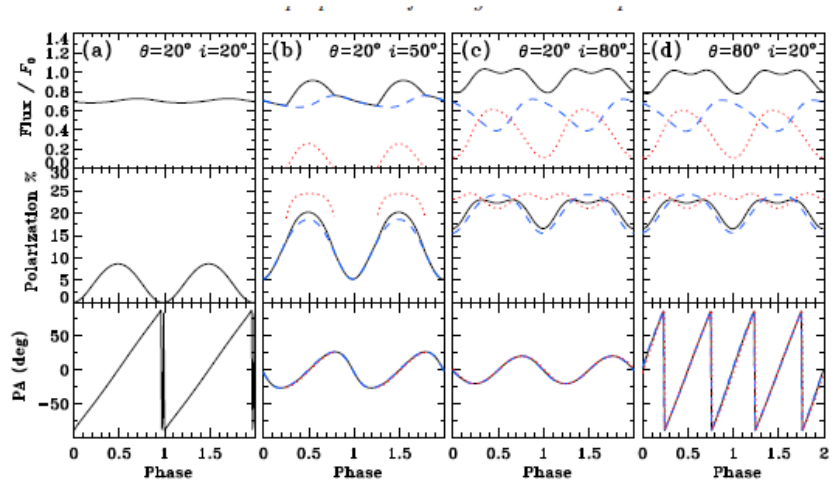}
\caption{Pulse profiles, polarization amounts and position angles for AMXPs. 
$\theta$ = inclination angle between rotation and magnetic axes, $i$ = 
observer angle referred to the rotation axis. From Poutanen 2010.\label{fig7}}
\end{figure}

Numerical calculations of the expected polarization properties of
Compton-scattered radiation emitted by such hot spots on neutron
stars have been done, including special relativistic boosting and 
aberration, and general relativistic light bending in Schwarzschild
space-time. For slowly-rotating neutron stars, the polarization
vector lies in the plane defined by the normal to the hot spot and
line of sight, as expected, since these are the basic directions in
a non-magnetic system. In fact, the polarization angle can be 
described by a rotating-vector type model (see Sec.\ref{sec:rpp})
which includes the special and general relativistic effects \citep{VP04}. 
Model pulse-phase profiles of $I$, $P_L$, and $\chi$ from such 
calculations are shown in Fig.\ref{fig7}. The idea in this case is also to
have an atlas of these profiles for a grid of values of the 
essential parameters like spot colatitude (referred to the rotation
axis of the neutron star) and observer angle, to be compared with
GEMS data on AMXPs. Such pulse phase polarimetry will be an excellent
diagnostic probe of the emission geometry in these systems.

\section{Low-Mass X-ray Binaries (LMXBs)}
\label{sec:lmxb}

Bright, accretion-powered galactic-bulge LMXBs are among the brightest 
X-ray binaries known, and are widely believed to be the nurseries
producing AMXPs by recycling. Their \emph{persistent} X-ray emission 
does not show periodicity at any possible rotation period relevant 
for neutron stars, although they 
are thought to harbor fast-rotating neutron stars with millisecond
periods. Various explanations have been offered for this, \eg, (a) that
the emitted pulses are \emph{``washed out''} by the dense plasma 
surrounding the neutron star, or (b) that their rotation and magnetic 
axes have become aligned by accretion torques, so that they \emph{actually} 
do not pulse. Thus, the kind of pulse phase polarimetry we have discussed 
so far in this paper is not possible for them.   

\subsection{Burst Oscillations}
\label{sec:burst}

However, many LMXBs undergo X-ray bursts, and coherent pulsations
at millisecond periods have been detected during these bursts in about
twenty of them so far. Identifying this periodicity with that of the neutron 
star's rotation, we can again use the above ideas of pulse phase polarimetry
as a diagnostic probe of the emission geometry in these systems. These 
bursts are thought to be due to the ignition of thermonuclear reactions in
the accreted material upon reaching a critical temperature and density,
deep in the atmosphere of the neutron star where the optical depth is
extremely large, so that the semi-infinite slab limit discussed in 
Sec.\ref{sec:amxp} applies. Furthermore, at sub-Eddington luminosities,
the burst process is not expected to ``lift'' the atmosphere to any significant
extent, so that the static slab limit still applies, and
straightforward calculations can and have been done with the aid of the 
Chandrasekhar-Sobolev formalism mentioned above.

\begin{figure}
\epsscale{0.9}
\plotone{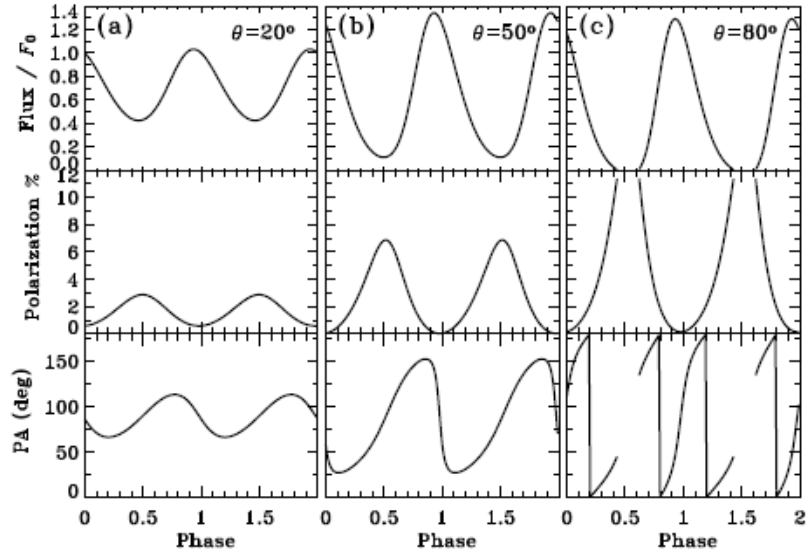}
\caption{Pulse profiles, polarization amounts and position angles for burst
oscillations from LMXBs. $\theta$ = colatitude of the ignition site and
$i$ = observer angle, both referred to the rotation axis. $i$ is taken
as 60 degrees in all cases shown. From Poutanen 2010.\label{fig8}}
\end{figure}

Results of such calculations \citep{VP04} are shown in Fig.\ref{fig8}, again detailing 
the model pulse-phase profiles of $I$, $P_L$, and $\chi$. The procedure here
is again to have an atlas of these profiles for a grid of values of the essential
parameters like the colatitude of the of the spot where the burst occurs,
and the observer angle, to be compared with GEMS data for diagnostics 
of the burst emission region.

\subsection{Quasiperiodic Oscillations (QPOs)}
\label{sec:qpo}

A major property of bright LMXBs in the time or frequency domain is the 
quasiperiodic oscillation (QPO) exhibited by them. QPOs appear as 
relatively wide, but clearly discernible, peaks in the power spectra of LMXBs,
as opposed to the extremely sharp peaks corresponding to periodic pulses
that appear in the power spectra of APPs. QPOs occur at both low 
frequencies ($\nu_{QPO}\sim 6-50$ Hz) and high ``kilohertz'' frequencies 
($\nu_{QPO}\sim 400 - 1000$ Hz), and they have also been detected recently
at intermediate ``hectohertz''frequencies \citep{vdk06}. The low-frequency (LF) QPOs
are particularly strong, and so were the first to be discovered \citep{vdketal85}. 
The study of QPOs and their relations with LMXB spectral states has become a 
most valuable diagnostic tool for probing the dynamics of inner accretion disks 
and the interactions between neutron stars and accretion disks. Studies of the 
polarization properties of QPO sources have a great potential for adding a new
dimension to this probe. 

Instead of the pulse phase polarimetry methods available for periodic sources, 
the appropriate diagnostic approach for QPO sources would be through 
studies of the cross-correlations between intensity and polarization parameters, 
and the associated cross-spectral analysis, which, in a sense, are the natural 
extensions of pulse-phase polarimetry to quasiperiodic sources. Schemes for 
such analysis are being constructed \citep{gs}.  The analog of the atlases for periodic
sources referred to in the earlier sections of this paper would, in this context,
be the atlases of cross-spectra for various QPO models, for a grid of values of
the essential parameters of each model.  Comparison of such atlases with the
observed cross-spectra obtained from GEMS data would discriminate between
models and constrain model parameters. As LF QPOs are particularly strong,
these would be the ideal first testing grounds for this scheme.

\section{Discussion} 
\label{sec:disc}

We have summarized here the principal polarization properties of the 
known classes of neutron-star X-ray sources, as expected from model
calculational schemes summarized in the earlier sections. Comparison
of the atlases of these expected polarization properties with GEMS
data will address some key scientific questions on X-ray emission
from neutron stars, which we now summarize. 

For APPs, the most immediate returns will be on the pencil/fan beam 
issue. Whereas beam-shape diagnostics from pulse-profile studies 
have been indirect and sometimes ambiguous, pulse phase polarimetry
provides a diagnostic, \viz, the phase-relation between the intensity
and the polarization degree or angle, which is simple, 
\emph{qualitatively} different for pencil and fan beams, and requires
only a coarse binning in pulse phase to achieve clear results.
However, the most fundamental discovery for APPs would be the 
discovery of the signature of vacuum resonance. As explained in 
Sec.\ref{sec:app}, the above phase-relation for a given beam-shape
changes sign at $E_v$. When $E_v$ lies in the GEMS band, even a very
coarse energy-binning will be able to find this signature.

For RPPs, the major scientific issue to be addressed is the nature
and site of the X-ray emission, as described by the various 
outer-magnetosphere models summarized in Sec.\ref{sec:rpp}. 
Together with the discrimination between these models through their
polarization signature, it may also be possible to obtain valuable
diagnostics of the magnetic field geometry, \eg, whether the field
is more similar to a retarded vacuum dipole (Deutsch field) 
configuration, or to a force-free configuration. 

For magnetars, the key issue is again the vacuum resonance point,
if it is indeed within the GEMS band for the observed magnetars.
In contrast to the phase-relation which changes sign at $E_v$
for APPs, it \emph{does not} do so for the primary radiation emitted
from the neutron-star atmosphere in magnetars \citep{LaiHo}. 
While the impact of propagation effects on this diagnostic feature 
need to be understood fully, all indications at the present time are that
they do not change it. This is thus a valuable signature of the 
superstrong field regime of magnetars, whose demonstration would
impact the fundamental physics (QED) of vacuum resonance. 

For AMXPs, the key subject amenable to polarizaton studies is the 
character of non-magnetic Compton scattering in neutron-star 
systems, for comparison with that of similar scattering from 
accretion disks is black-hole systems, particularly at those 
radii in the latter systems where GR effects are mild. This bears
on the important issue of the possible impact of the remnant 
magnetic field in AMXPs on the scattering and polarization 
signature. This magnetic field is still invoked for envisaging
magnetic chanelling and polar hot-spots of X-ray emission, and
indeed one of the model parameters obtained by comparing the 
atlases with GEMS data would be the angle between the magnetic
and rotation axes. Any residual signature of this magnetic 
field on X-ray polarization is therefore an intriguing 
possibilty. 

For LMXB burst oscillations, a key question is the site of 
thermonuclear ignition and X-ray bursts on the neutron-star
surface. Recent observations suggest various possible sites 
between the rotational equator and pole for the onset of ignition.  
Polarization diagnostics would give us the angular position 
of this site in relation to the rotation axis, thus clarifying how
an X-ray burst develops as the burning front spreads on the
rotating star. Once again, the role of the residual magnetic 
field in LMXBs remains an interesting issue here.

Finally, a key target of polarization studies in LMXBs with QPOs 
woud be to discriminate between various QPO models, particularly
those aspects of the models which deal with the interaction of the 
accretion disks in these systems with the remnant magnetic field in
neutron stars in LMXBs. The role of rotating clumps or blobs of
plasma in quasiperiodic X-ray emission has been studied intensively
in this context, and polarization diagnostics will add a new dimension
to these studies.

\section{Epilogue} 
\label{sec:epi}

GEMS was proposed in response to a NASA SMEX announcement of
opportunity in December 2008. It was selected for phase A development 
in 2009 and down-selected for phase B in 2010. A technically
successful Preliminary Design Review was held in Feb 2012. NASA
Science Mission Direectorate (SMD) indicated their intention to 
non-confirm (or cancel) in May 2012; the SMD decision was based on
concerns that the eventual cost would be too high.

\clearpage
%
%
%

\clearpage

\clearpage

\clearpage

\clearpage

\clearpage

\clearpage

\clearpage

\clearpage

\clearpage
 
\clearpage
 
\end{document}